# Slowing Down of Ring Polymer Diffusion Caused by Inter-Ring Threading[a]

Eunsang Lee, Soree Kim and YounJoon Jung*

---

Department of Chemistry, Seoul National University, Seoul 151-747, Korea

E-mail: yjjung@snu.ac.kr

---

Diffusion of long ring polymers in a melt is much slower than the reorganization of their internal structures. While direct evidences for entanglements have not been observed in the long ring polymers unlike linear polymer melts, threading between the rings is suspected to be the main reason for slowing down of ring polymer diffusion. It is, however, difficult to define the threading configuration between two rings because the rings have no chain end. In this work, evidences for threading dynamics of ring polymers are presented by using molecular dynamics simulation and applying a novel analysis method. The simulation results are analyzed in terms of the statistics of persistence and exchange times that have proved useful in studying heterogeneous dynamics of glassy systems. We find that the threading time of ring polymer melts increases more rapidly with the degree of polymerization than that of linear polymer melts. This indicates that threaded ring polymers cannot diffuse until unthreading event occurs, which results in the slowing down of ring polymer diffusion.

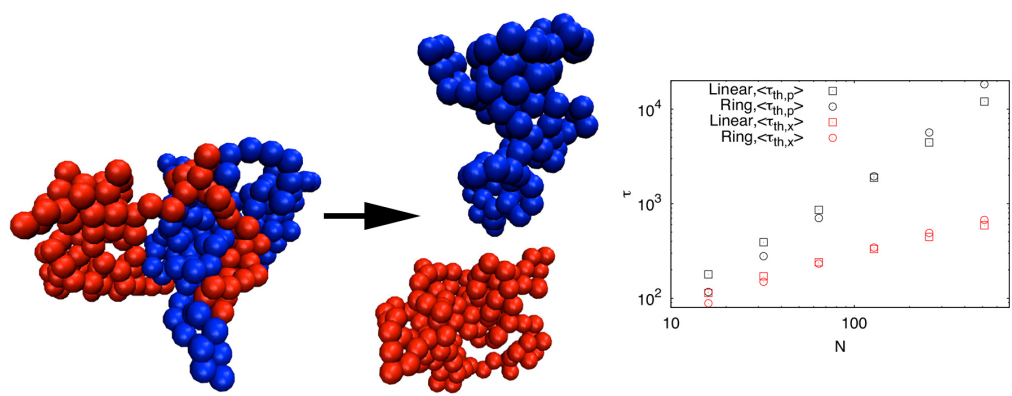

# 1. Introduction

Rheological properties of polymeric materials have been of great interest to many researchers in recent years. In general, polymers show different dynamical behaviors with their inherent topologies. For example, linear and branched polymers show different dynamical scaling behaviors.[1,2] Recently, ring polymers have brought about novel issues to the dynamics in polymeric melt phase. Their crumpled globular structures with high fractal dimensions on a sufficiently large length scale resemble several biological molecules, existing as chromosomal territories in cells.[3-8] One of the interesting dynamical features is the diffusion behavior of unknotted, unlinked ring polymers. A few years ago, Kawaguchi and coworkers succeeded in synthesizing highly mono-disperse ring polymers,[9,10] and found that the diffusive motion of ring polymer melts rapidly slows down as the degree of polymerization, $N$, increases. In this case, it was found that the diffusion coefficient shows a scaling behavior as $D \sim N^{-1.9}$. A number of simulation studies also reported similar scaling relations of the diffusion coefficient with its scaling exponent ranging from -2.4 to -1.9.[11-14] Interestingly, it turns out that these values are very similar to those found in the reptation model of linear polymer melts, although topological constraints in ring polymers do not allow them to move around as in the reptation model. It is also noteworthy that the scaling relations are consistent with a few theoretical works based on the lattice animal model of the double-stranded conformation of the rings.[15,16] However, a plateau modulus that is characteristics of entangled polymer melts, and systematic double-stranded configurations have not been observed in the ring polymer simulations.[11] These observations present intriguing questions regarding the nature of diffusive behaviors in the ring polymer melts.

A couple of simulation studies have provided some clues to the mechanism of diffusive motions in ring polymer melts. One of them is decoupled time scales between diffusion and structure reorganization.[11] This decoupling behavior has not been observed in entangled linear polymers, whose scaling exponent of the disentanglement time versus $N$, which turns out to be 3, is exactly the same as that of the structure relaxation time.[17] The fact that the ring polymer diffusion time is much larger and more rapidly increases with $N$ than the internal structure relaxation time has been used as an evidence to support the idea of amoeba-like motion of ring globules.[11,18]

It is fair to say that neither the simulation nor the theory has succeeded in suggesting a clear mechanism of ring polymer dynamics yet. One of the difficulties in simulation studies is defining a threading configuration in the melt phase. One of the most well established and generally applied analysis methods in revealing entanglement effects is primitive path analysis (PPA) pioneered by Kremer and coworkers.[19,20] In linear polymer melts, the PPA can provide the shortest path of polymers by fixing positions of both end monomers, and can reveal the entanglement effect between different chains, which then allows one to calculate the entanglement length and the plateau modulus. Halverson et al. applied PPA to ring polymers by fixing two opposite particles randomly, and clearly demonstrated entangled rings, although they were not able to identify clearly the effects of entanglement on the dynamics.[11] We also note in passing that Michieletto et al.[21,22] studied slowing down of the ring polymers in a pre-constructed gel structure. Although these works do not give direct evidence for the threading effect on the dynamics of ring polymer melts, it provides an intriguing question for the threading effect on the ring polymer dynamics.

In this work, we provide an alternative measure for calculating the threading time, that is, statistical distributions of so-called persistence and exchange times of molecular contacts in polymer melts. The distributions of persistence and exchange times and their decoupling behaviors have been originally developed in the study of dynamical heterogeneity observed in glassy systems, and have been used very successfully as a measure to characterize fluctuation-dominated dynamics in supercooled systems and ionic liquid systems.[23-26] We utilize this new method to study how threading affects dynamical slow-down of ring polymer diffusions by studying statistical distributions of persistence times in molecular contacts.

This paper is organized as follows. In section 2, we first briefly explain the model and the simulation methods. In section 3, we provide a definition of threading time, and show how the threading motions affect the dynamical behavior of ring polymers. Comparisons between ring threading with linear entanglement also provide molecular evidences for elucidating slow-down of ring polymer dynamics. Concluding remarks follow in the final section.

## 2. Simulation Method

We perform molecular dynamics simulations using a bead-spring model developed by Kremer and Grest.[17] In this model, potential energy functions for non-bonded interactions, bond stretchings, and angle bendings are represented by Weeks-Chandler-Andersen (WCA), Finite-Extensible-Nonlinear-Elastic (FENE), and cosine angle potentials, respectively. In particular, FENE potential prevents different chains from crossing each other. A weak temperature coupling method is applied to the system using the Langevin thermostat. We simulate linear and ring polymer systems for various cases with different number of monomers in a polymer, $N$ = 16, 32, 64, 128, 256, and 512. All systems consist of 128 monodisperse polymer molecules and the number density of monomers is set to be 0.85. While the entanglement length of linear polymers for this model is known as $N_{e,linear} \approx 28$,[19] that of a ring has not been well-established yet. The recent numerical study reported $N_{e,ring} \approx 77$ for rings, by counting the number of monomers in a Kuhn segment of the primitive path in analogy to the linear polymer,[11] but interplay between $N_{e,ring}$ and the ring dynamics still remains unclear. Although our longest ring does not reach an asymptotic scaling regime in their sizes,[27] its length is much longer than $N_{e,ring}$ and it has a value of Flory exponent, $v$ = 2/5, which is smaller than that of short rings and linear polymers. Thus it is expected that our long rings are long enough to study threading effects in ring polymers. After equilibrating the system during time longer than three times of the Rouse time,[2,28] we perform production runs until the mean square displacement of polymer's center-of-mass motion reaches a diffusive regime.

As a control system, we also perform simulation studies of diffusional motions of spherical particles, which mimic the dynamics of a very compact globular ring polymer without a threading effect. The radius of the sphere is chosen such that the radius of gyration and the mass of the sphere are the same as those of a polymer. We set the volume fraction to be the same with the polymer system. The other simulation parameters are the same as those of polymer systems. GROMACS package is used in all equilibration and production runs.[29] Supporting information is available for more simulation details.

## 3. Results and Discussion

A typical threading configuration between two rings is shown in **Figure 1** in terms of a schematic representation (a and b) and a simulation snapshot for $N$ = 128 ring system (c and d). In Figure 1(a) and (b), the threaded ring (red) cannot diffuse or diffuse together with the treading ring as long as the threading configuration is maintained. When the threading ring (blue) unthreads as in Figure 1(b), the threaded one can finally diffuse away. This picture is also illustrated by simulation snapshots shown in Figure 1(c) and (d), where a red ring forming a large loop is threaded by a blue polymer. These figures illustrate the basic idea of threading dynamics of ring polymers, that is, why rings diffuse so slowly even through there is no entanglement effect.

One of the important observations in threading dynamics is that diffusion time and structure relaxation time of ring polymers are decoupled, unlike those of linear chains. **Figure 2** shows the diffusion and structure relaxation times of linear and ring polymers. Diffusion time is defined by $\tau_d = <R_g^2>/D_G$, where $D_G$ is diffusion coefficient of center of mass, and structure relaxation time is calculated by $\tau_{sr} = \int_0^\infty C(t)dt$, where $C(t)$ is a time auto-correlation function of end-to-end vectors for linear polymers, and spanning vectors for ring polymers (More details are given in Supporting information). This figure clearly shows that, in the linear polymer case, both time scales are almost the same, and they are scaled by $\tau \sim N^3$. However, in the ring polymer case, not only the absolute values of diffusion times but the scaling exponent versus $N$ are much larger than those of structure relaxation times, which means that the rings reorganize their internal structures before they move around about the distance corresponding to their sizes. This result is also consistent with the previous study of Halverson et. al., which shows the decoupling of diffusion coefficients and friction coefficients for ring polymer melts.[11] Fast memory loss of spanning vectors is mainly caused by spinning motions of rings around the specific axis formed by the threading polymer without diffusion (Figure 1(a)), which can be a good evidence for threading dynamics of ring polymers.

For analyzing entangled configurations, the PPA method has been used for linear polymer melts.[19,20] However, it is not straightforward to extend the method to the ring polymer melts. The PPA may also be likely to miss some of the threading configurations for ring polymers when randomly picked two monomers of a threading polymer are located along the same direction from the needle plane formed by a loop of the threaded polymer. Therefore, in this paper, we propose an alternative method for obtaining statistics of threading times based on persistence time distributions of molecular contacts. Persistence times and related exchange times were first proposed by Jung et al.[23] as a good measure to identify dynamically heterogeneous behavior of glassy systems. In glass forming liquids that are typically modeled by atomistic Lennard-Jones liquids or kinetically constrained lattice models, mobile and immobile regions of the system are identified and labelled with a binary variable, $n_i$. Usually, the binary variable is assigned 1 or 0, whether it corresponds to a mobile or immobile region, respectively. In such coarse-grained models, persistence time, $\tau_p$ is defined by the time duration over which a mobile (or an immobile) region persists from $t = 0$ until $t = \tau_p$, where the state of the mobility changes. A related, but different statistical measure is called an exchange time, $\tau_x$, as a time duration between two mobility-changing (or exchange) events at time $t$ and time $t + \tau_x$.[23] For stochastic model systems, it is known that there is an exact relation between probability distributions of persistence ($P_p(t)$) and exchange times ($P_x(t)$), such that[23]

$$P_p(t) = \frac{\int_t^\infty dt' P_x(t')}{<\tau_x>}, \qquad (1)$$

where $<\tau_x> = \int_0^\infty dt\, t P_x(t)$ is a mean exchange time. Moreover, for Poisson processes where exchange events are not correlated to each other, two probability distributions are identical to each other and they have exponential decay forms in time. However, when there are strong correlations between the exchange events, the two distributions decouple from each other.

To adjust this idea to polymer systems, we characterized molecular contact between two polymers as a binary variable. To do so, we first define $s(i, j; t)$ as a binary, contact variable, which takes 1 when the centers of mass of the $i$-th and $j$-th polymers are separated less than the sum of their radii of gyrations at time $t$, and 0 otherwise,

$$s(i,j;t) = \begin{cases} 1, & \text{if } |\vec{r}_{CM,i}(t) - \vec{r}_{CM,j}(t)| < R_{g,i}(t) + R_{g,j}(t) \\ 0, & \text{otherwise} \end{cases}, \qquad (2)$$

where $\vec{r}_{CM,i}(t)$ and $R_{g,i}(t)$ are the center of mass position and the radius of gyration of $i$-th polymers at time $t$, respectively. If one ring threads another, those two rings are expected to be in a contact with each other. Therefore, using the contact variable, we can study dynamical behavior of the threading events by calculating probability distributions of persistence and exchange times for contact pairs. Since threading configurations only contribute to the contact pairs, we count persistence and exchange times for the contact pairs only.

We first tested this method by applying the idea to a simple system consisting of spherical particles. In this case we used the radius of the sphere when calculating contact variables in Equation 2. **Figure 3** shows probability distributions of persistence and exchange times of (a) spherical system corresponding $N = 32$ ring polymers, (c, top) linear polymers, and (c, bottom) ring polymers. In the case of spherical particles, one can see that the probability distributions of persistence and exchange times are exponential and identical to each other as shown in Figure 3(a) and its inset. This indicates that the dynamics of spherical particles is completely Poissonian, caused by random diffusional motions.

In polymeric systems, however, there are two different time scales that contribute to statistics of persistence and exchange time distributions. The first one is random diffusion which is characterized by an exponential distribution at short time as in the simple sphere system. The more important and interesting one is long persistence and exchange times shown as long tails in the distributions of Figure 3(c). Because of this long tail, two distributions of polymer systems are decoupled from each other especially for long polymers. According to the relation between persistence and exchange times given in Equation 1, higher probability of finding longer exchange times results in fat tails in the distribution of persistence times at larger values. It is an intriguing question to ask what mainly contributes to long persistence and exchange times. As mentioned before, we expect that if one ring is threaded by another ring, a contact is formed between them. For linear polymers, molecular contacts are well captured by entanglement effects, because one polymer is confined in loops formed by another polymer. For ring polymers, however, diffusional motions of the threaded ring is restricted due to the threading ring, which prevents them from separating out. Therefore, we expect that the threading dynamics of ring polymer melts is mainly responsible for long exchange and persistence times.

Following the idea presented above, we calculated distributions of threading times. Figure 3(b) shows linear-log plots of persistence and exchange time distributions (open squares and circles, respectively) for $N = 512$ ring polymers. In a very short time, the probability distributions decay almost exponentially as shown by the linear plots in log-linear scales. Since this short time part is due to the diffusion of non-threading polymers, not threading dynamics, we subtract this part from the original distributions after fitting the short time distribution to an exponential function, $P_{short}(t) = c_s \cdot \exp(-t/\tau_s)$, and then renormalize the distributions. The resulting distributions of threading times are shown as filled squares and circles in Figure 3(d), and they look similar to each other. In the case of linear polymers, it takes some time, called disentanglement time, for an entangled polymer to escape from the previous confining tube and to be located in a new tube. This disentanglement time is closely related to threading time which it takes for two threaded polymers to separate out. Therefore, as in the case of entangled linear polymers, a threaded ring polymer also needs unthreading time to escape from threading configuration. The fact that threading time of ring polymer is related to disentanglement time of linear polymers makes it possible to anticipate threading effect on ring polymer dynamics.

Lastly, to quantitatively compare the threading times of linear and ring polymers, we obtained average persistence and exchange times for threading configurations. In **Figure 4**, it is shown that, for polymers of $N < 64$, both threading times of ring polymers are smaller than those of linear ones, but for long polymers of $N > 128$, threading times of rings dominate over those of linear polymers. Interestingly, the crossover length between $N = 64$ and 128 turns out to be similar with the entanglement length of ring polymers, $N_{e,ring}$=77, as revealed in a previous study using PPA.[11] It is also interesting to note that the crossover length scale is related to the universal behavior of polymer sizes. According to a number of previous researches studying ring polymer sizes,[5,6,11,30] Flory exponent decreases from 1/2 to 1/3 depending on degree of polymerization. The crossovers between different scaling regimes are universal, independent of ring polymers simulated by different models.[27] The first crossover from $\nu$ = 1/2 to 2/5 is at $N \approx 3N_{e,linear}$ and the second from 2/5 to 1/3 is at $N \approx 30N_{e,linear}$, which is a characteristic feature of ring polymer melts. However, linear polymer melts have a Flory exponent of 1/2, regardless of the degree of polymerization.

From a theoretical point of view, threading time is closely related to the self-excluded monomer volume. For a crumpled globule, it is well known that a relatively small fraction of inner volume of a ring polymer is penetrated by surrounding polymers.[5] This small self-excluded volume contributes to the threading dynamics in two different ways. The first contribution comes from the reduced inter-ring overlap which leads to a small threading probability. The second is, in contrast, that an elongated contour length of the threaded primitive path increases threading times. In analogy to the reptation model of entangled linear polymer melts, the threading polymer squeezes through the loops of surrounding threaded polymers. The squeezing time which is regarded as a threading time is proportional to the contour length of the primitive chain for the threaded polymer, and this contour length is also monotonically increases when the inner space for surrounding rings to interpenetrate diminishes. Our observation on the threading times indicates that the latter effect is a dominant one in determining an average threading time. Even though the long ring in a melt has a low probability of threading, once the threading is made, a ring spends much time to unthread, which results in the slowing down of the ring dynamics. The fact that the degree of polymerization at the first crossover is approximately the same with our crossover length of average threading times between linear and ring polymers strongly supports this argument. The polymer length corresponding to the second crossover, $N \approx 840$, is out of range in our simulation studies. However, we expect that more compact globular rings with a small self-excluded monomer volume exhibit very long threading times, which provides a major contribution to the slowing down of ring polymer dynamics.

## 4. Conclusion

In this work, we show that threading and unthreading events are important factors in determining dynamics of ring polymers by using persistence and exchange times of polymer contacts. As opposed to linear polymer melts, Flory exponent of ring polymer melts decreases from 1/2 to 1/3 as degree of polymerization increases. Since the small Flory exponent of polymers leads to a small self-excluded monomer volume, ring polymers take longer time to squeeze from the threading configuration than linear polymer melts. If one ring threads another, diffusion of the threaded ring is restricted and the overall diffusion of ring polymer melts slows down. Slow diffusion of the threaded ring is represented by a fat tail in the exchange time distribution of molecular contacts, resulting in decoupled persistence and exchange time distributions. The degree of polymerization at which the threading time becomes important is consistent both with the entanglement length of ring polymers, and the crossover length of Flory exponent.

In spite of the successful analysis of the threading effect of ring polymers presented above, we note that the average threading times are about ten times smaller than the structure relaxation times and diffusion times. The exponent in scaling behavior of average threading time versus degree of polymerization, that is $\tau_{\text{th,p}} \sim N^2$ in Figure 4, is also somewhat smaller compared to diffusion time. It may be the case that there are some threading configurations that our definition of molecular contact cannot capture. For example, threading configurations of two small loops sticking out of their main bodies in which the distance between their centers of mass is longer than the sum of their radii of gyrations is not recognized by the contact. To understand the effects of the missing threading configurations on the overall distribution of persistence and exchange time distributions, a more precise definition of threading configurations may be necessary. It is also desirable to develop an analytical theory of treating threading configurations using static properties of polymers, including looping probability, contact probability between molecules, and the size of sub-chains to help further understanding of threading dynamics of longer ring systems that are not easily studied by molecular dynamics simulations. Various aspects of threading dynamics along these lines are currently under study.

## Supporting Information

Supporting information is available from the Wiley Online Library or from the author.

Acknowledgements: This work is supported by the IRTG program (No. 2011-0032203) of National Research Foundation of Korea. The authors acknowledge the financial support from the National Research Foundation of Korea (Grant Nos. NRF-2012R1A1A2042062 and NRF-

2007-0056095), and the Brain Korea 21 Plus Program. Computational resources have been provided by KISTI supercomputing center through Grant No. KSC-2013-C2-021.

Keywords: ring polymers; diffusion; threading dynamics; persistence and exchange time; scaling

Figures

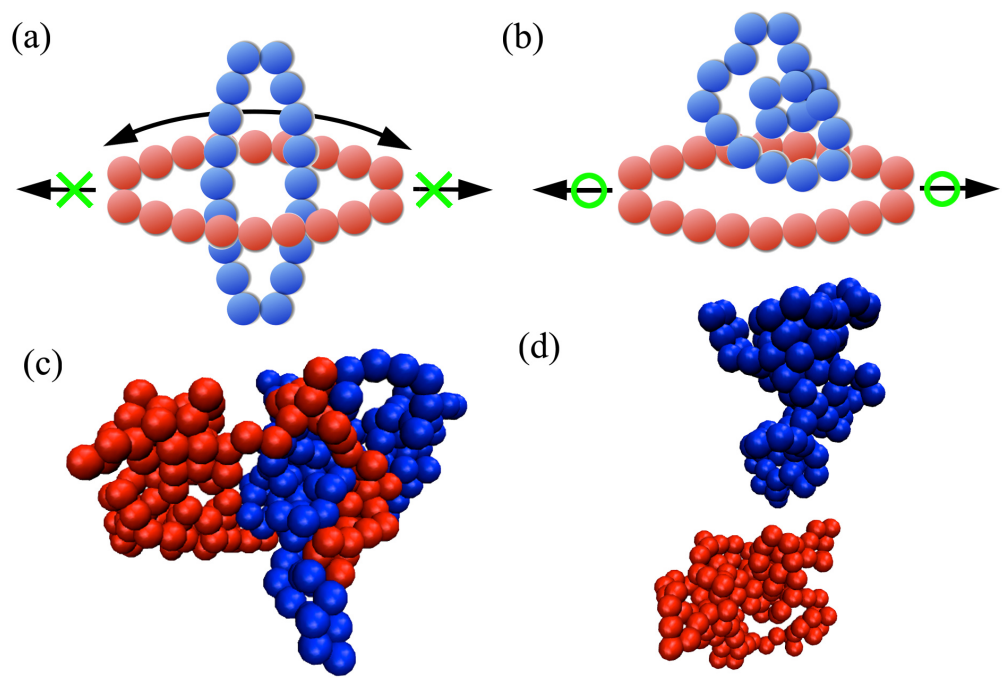

*Figure* 1. (a, b) Schematic descriptions of threading dynamics and (c, d) simulation snapshots of threading configuration.

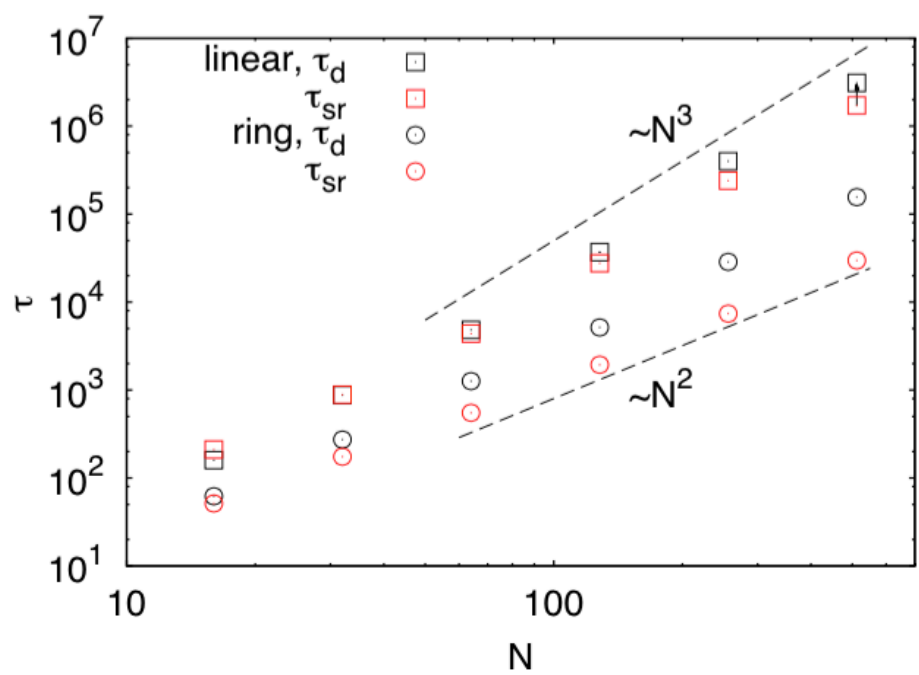

*Figure* 2. Diffusion time (black, $\tau_d$) and structure relaxation times (red, $\tau_{sr}$) of linear (squares) and ring (circles) polymers.

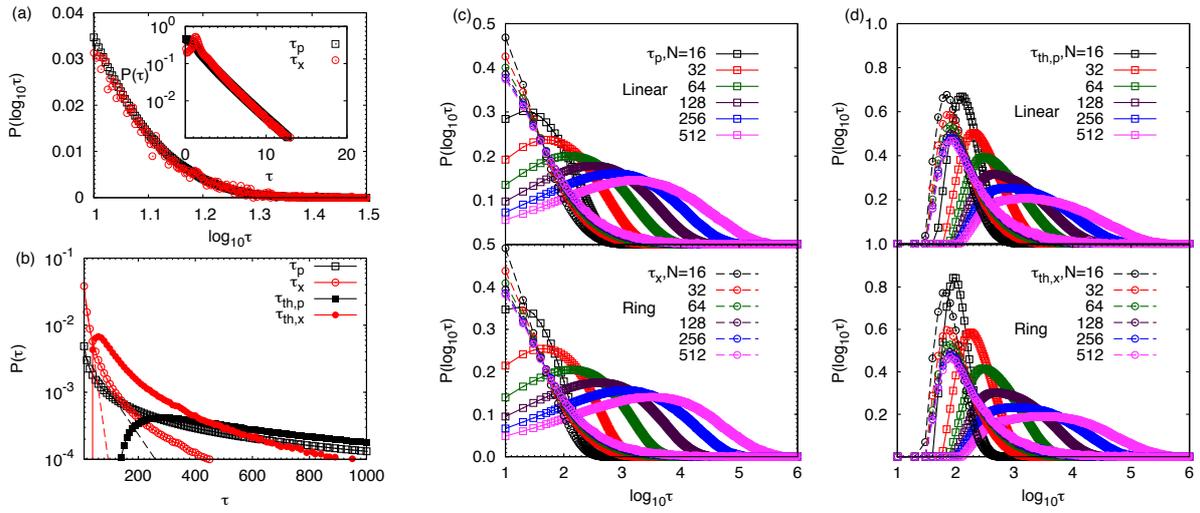

*Figure* 3. (a) Probability distributions of persistence (black squares) and exchange (red circles) times for a sphere system and (c) those for linear (top) and ring (bottom) polymers. All figures are represented by $P(\log_{10}\tau)$ versus $\log_{10}\tau$ except for the inset of (a) which shows identical exponential decay of persistence and exchange time distributions. Figure (b) is linear-log plot of original persistence and exchange time distributions (open squares and circles) of $N = 512$ ring polymers with fitted exponential functions at short time shown by dashed lines. Threading time distributions are plotted by filled squares and circles. Figure (d) shows probability distributions of threading times.

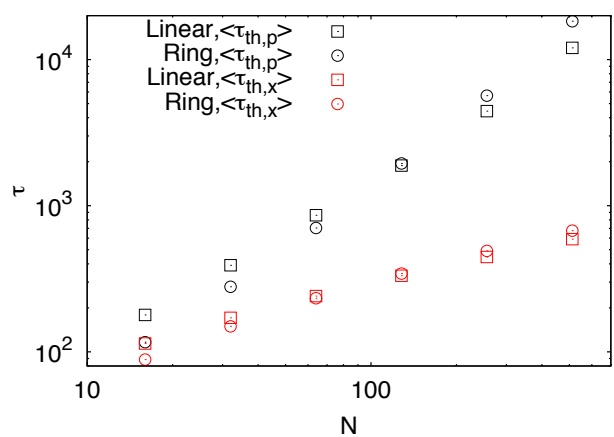

*Figure* 4. Average threading persistence (black) and exchange (red) times of linear (squares) and ring (circles) polymers. The sizes of error bars for all systems are much smaller than the size of symbols. Details are given in Supporting Information.

# Supporting Information

## 1. Simulation Details

In the Kremer-Grest bead-spring model, a polymer consists of beads with mass $m$. Nonbonded beads interact with each other through the repulsive Lennard-Jones (Weeks-Chandler-Anderson) potential (Equation S1). The covalent bond between beads is described by the Finite Extensible Nonlinear Elastic (FENE) potential (Equation S2), and the cosine bond angle potential (Equation S3) is used to model semi-flexible polymers.

$$U_{\text{pair}}(r_{ij}) = \begin{cases} 4\varepsilon\left[\left(\frac{\sigma}{r_{ij}}\right)^{12} - \left(\frac{\sigma}{r_{ij}}\right)^{6} + \frac{1}{4}\right], & r_{ij} < 2^{1/6}\sigma \\ 0, & \text{otherwise} \end{cases} \quad (S1)$$

$$U_{\text{bond}}(r_{ij}) = \begin{cases} -0.5kR_0^2 \ln\left[1 - \left(\frac{r_{ij}}{R_0}\right)^2\right], & r_{ij} < 2^{1/6}\sigma \\ \infty, & \text{otherwise} \end{cases} \quad (S2)$$

$$U_{\text{angle}}(\theta_i) = k_\theta[1 - \cos(\theta_i - \pi)]. \quad (S3)$$

$K = 30\varepsilon/\sigma^2$, $R_0 = 1.5\sigma$ and $k_\theta = 1.5\varepsilon$ are used in the above equations. The unit time is scaled by $\tau = \sigma\sqrt{m/\varepsilon}$. For all simulations, we use the weak temperature coupling methods, that is, Langevin thermostat. Obviously, periodic boundary condition is used to all directions. In this study, we focus on the properties of unlinked, unknotted ring polymers and such configurations are prepared as follows. At first, we place monomers with perfect circular morphology, such that planes of the circles are perpendicular to $z$-axis. Centers of the circles are placed on square lattice sites on the $xy$-plane. To avoid concatenation between the polymers, the lattice constant of the square lattice should be larger than the diameter of a fully extended circle, so we set the lattice constant to $2.5r = 2.5N2^{1/6}\sigma/2\pi$, because $N2^{1/6}\sigma \approx 2\pi r$ for $N \gg 1$, where $r$ is a radius of a circle. At high pressure of $P = 5.0\varepsilon/k_B$, short NPT simulations are performed until desired values of the system size satisfying $\rho = 0.85\sigma^{-3}$ are obtained. For all systems, desired densities are obtained within $5\times 10^3 \tau$.

## 2. Mean Square Displacement and Diffusion Coefficient

In simulation studies of polymers, a different types of the mean square displacements (MSDs) are calculated,

$$g_1(t) = <[\vec{r}_i(t) - \vec{r}_i(0)]^2> \quad (S4)$$

$$g_2(t) = <[\vec{r}_i(t) - \vec{r}_{CM}(t) - \vec{r}_i(0) + \vec{r}_{CM}(0)]^2> \quad (S5)$$

$$g_3(t) = <[\vec{r}_{CM}(t) - \vec{r}_{CM}(0)]^2> \quad (S6)$$

Here, $\vec{r}_i$ and $\vec{r}_{CM}$ are the positions of a monomer and a center of mass, respectively. $g_1(t)$ is the ensemble average of the monomer displacement and $g_3(t)$ is that of the center-of-mass displacement. $g_2(t)$ is a kind of time-correlation function of the distance between a monomer and the center-of-mass, which starts from 0 at $t = 0$ and converges to $<R_g^2>$ at long time.

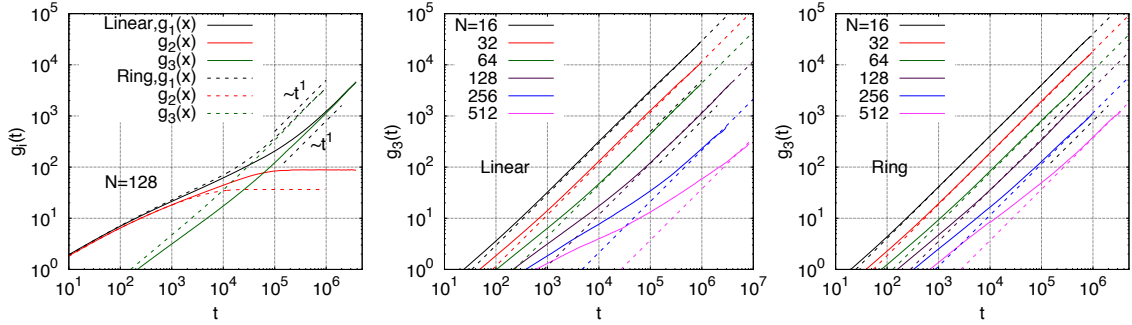

*Figure* S1. (a) $g_1(t)$, $g_2(t)$, and $g_3(t)$ for linear (solid lines) and ring (dashed lines) polymers. $g_3(t)$ for (b) linear and (c) ring polymer melts are also shown.

**Figure S1** shows MSDs of $N = 128$ linear and ring polymer melts. At long time, all the systems except for $N = 512$ linear polymers reach diffusive regime, at which $g_3(t)$ is proportional to $t^1$. We obtain diffusion coefficients, $D_G$, by fitting the diffusive regime of $g_3(t)$ to the equation $g_3(t) = 6D_G t$.

## 3. Structure Relaxation Time

Time-correlation functions of the end-to-end vectors for the linear polymers, and spanning vectors for the ring polymers are used to calculate the structure relaxation times,

$$C(t) = <\vec{r}_{ee}(0) \cdot \vec{r}_{ee}(t)>, \quad (S7)$$

where

$$\vec{r}_{ee}(t) = \begin{cases} \vec{r}_1(t) - \vec{r}_N(t), & \text{for linear polymers,} \\ \vec{r}_i(t) - \vec{r}_{i+N/2}(t) & \text{for ring polymers.} \end{cases} \quad (S8)$$

More precisely, when calculating the positions of $i$-th monomer in a ring polymer, we average the positions of five adjacent monomers from ($i$-2)-th to ($i$+2)-th monomers. **Figure S2** shows the time-correlation function of the linear and ring polymers, and they show stretched exponential behaviors. We can obtain the structure relaxation times by integrating the correlation function from 0 to infinity.

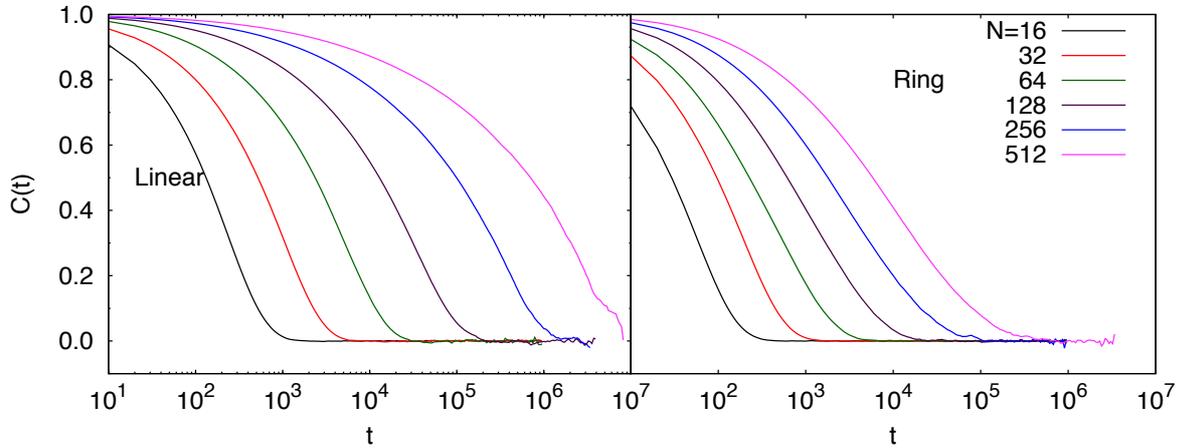

*Figure* S2. Structure relaxation times of (a) linear and (b) ring polymers. Different polymer lengths are depicted in different colors.

## 4. Short Time Fitting of Persistence and Exchange Time Distributions

In order to subtract the short time contribution of Poissonian diffusion from persistence and exchange time distributions, we fit the short time distributions by an exponential function, $P_{\text{short}}(t) = c_s \cdot \exp(-t/\tau_s)$. Fitted parameters, $c$ and $\tau$ are described in **Table S1**. Errors caused by fitting range are negligible. Then, we subtract this short time contribution from the overall distributions to get the distributions of threading times only.

*Table* S1. Short time fitting parameters of persistence and exchange time distributions to the function, $P_{\text{short}}(t)$. Subscripts, "p" and "x" mean the parameters corresponding to the persistence and exchange time distributions respectively.

| N | Linear | | | | Ring | | | |
|---|---|---|---|---|---|---|---|---|
|  | $c_{s,p}$ | $\tau_{s,p}$ | $c_{s,x}$ | $\tau_{s,x}$ | $c_{s,p}$ | $\tau_{s,p}$ | $c_{s,x}$ | $\tau_{s,x}$ |
| 16 | 0.369 | 24.594 | 1.016 | 12.617 | 0.531 | 20.937 | 1.097 | 12.183 |
| 32 | 0.225 | 34.560 | 0.863 | 13.685 | 0.259 | 32.406 | 0.914 | 13.174 |
| 64 | 0.140 | 45.273 | 0.784 | 14.285 | 0.146 | 44.090 | 0.821 | 13.731 |
| 129 | 0.092 | 56.149 | 0.743 | 14.582 | 0.090 | 54.927 | 0.779 | 14.101 |
| 256 | 0.047 | 65.179 | 0.723 | 14.709 | 0.059 | 63.918 | 0.756 | 14.260 |
| 512 | 0.047 | 71.530 | 0.716 | 14.710 | 0.042 | 72.677 | 0.750 | 14.263 |

## 5. Errors of the Average Threading Times

The threading time distributions are obtained and renormalized after subtracting the short time contribution as explained in the main text in details. Errors of the average threading times in Figure 4 can be estimated by those of contact persistence and exchange times listed in the Table S2, in which the standard errors(SE) are about four orders of magnitude smaller than its average values for all systems. Since we use only short time data set when fitting distributions, errors caused by the fitting procedure are also negligible as shown in Figure 3(b). Therefore, we expect that the standard errors of the average threading times are much smaller than the symbol size in Figure 4.

*Table* S2. Averages and standard errors of the contact persistence and exchange times.

| $N$ | Linear | | | |
|---|---|---|---|---|
| | $\langle\tau_p\rangle$ | $SE(\langle\tau_p\rangle)$ | $\langle\tau_x\rangle$ | $SE(\langle\tau_x\rangle)$ |
| 16 | $6.5139\times10^0$ | $1.0752\times10^{-3}$ | $3.5189\times10^0$ | $6.5941\times10^{-4}$ |
| 32 | $1.6563\times10^1$ | $2.8035\times10^{-3}$ | $5.2092\times10^0$ | $1.1642\times10^{-3}$ |
| 64 | $4.2377\times10^1$ | $7.0837\times10^{-3}$ | $7.4175\times10^0$ | $1.9826\times10^{-3}$ |
| 128 | $1.1012\times10^2$ | $1.7916\times10^{-2}$ | $1.0335\times10^1$ | $3.3756\times10^{-3}$ |
| 256 | $2.8450\times10^2$ | $3.8029\times10^{-2}$ | $1.3853\times10^1$ | $4.8201\times10^{-3}$ |
| 512 | $8.5024\times10^3$ | $3.7516\times10^{-1}$ | $1.8303\times10^2$ | $9.2414\times10^{-3}$ |
| $N$ | Ring | | | |
| | $\langle\tau_p\rangle$ | $SE(\langle\tau_p\rangle)$ | $\langle\tau_x\rangle$ | $SE(\langle\tau_x\rangle)$ |
| 16 | $4.0350\times10^0$ | $8.4147\times10^{-4}$ | $2.8836\times10^0$ | $6.5449\times10^{-4}$ |
| 32 | $1.1291\times10^1$ | $2.5664\times10^{-3}$ | $4.6585\times10^0$ | $1.3615\times10^{-3}$ |
| 64 | $3.4681\times10^1$ | $8.2982\times10^{-3}$ | $7.1503\times10^0$ | $2.7674\times10^{-3}$ |
| 128 | $1.1434\times10^2$ | $2.8169\times10^{-2}$ | $1.0542\times10^1$ | $5.6263\times10^{-3}$ |
| 256 | $3.8270\times10^2$ | $9.8045\times10^{-2}$ | $1.4973\times10^1$ | $1.1315\times10^{-2}$ |
| 512 | $1.3563\times10^4$ | $3.3556\times10^{-1}$ | $2.0485\times10^2$ | $1.8805\times10^{-2}$ |